\documentclass[onecolumn,superscriptaddress,preprintnumbers,nofootinbib,notitlepage]{revtex4}

\usepackage{bm}
\usepackage{feynmp}
\usepackage{simplewick}
\usepackage{graphics,graphicx}
\usepackage{amsmath,amssymb}
\usepackage{color}
\usepackage{array,mathtools}
\usepackage{booktabs,multirow}
\usepackage{bbold,mathrsfs}
\usepackage[normalem]{ulem}
\usepackage[colorinlistoftodos,prependcaption,textsize=tiny]{todonotes}		

\newcommand{\bea}{\begin{eqnarray}}
\newcommand{\eea}{\end{eqnarray}}

\newcommand{\beq}{\begin{equation}}

\newcommand{\eeq}{\end{equation}}

\newcommand{\MeV}{{\rm MeV}}
\DeclarePairedDelimiterX\brakket[3]{\langle}{\rangle}{#1\,\delimsize\vert\,#2\,\delimsize\vert\,#3}%
\DeclarePairedDelimiterX\braket[2]{\langle}{\rangle}{#1\,\delimsize\vert\,#2}%
\DeclarePairedDelimiterX\braet[2]{\langle}{\rangle}{\,#1\,#2\,}%
\DeclarePairedDelimiterX\bra[1]{\langle}{\rvert}{#1}%
\DeclarePairedDelimiterX\ket[1]{\lvert}{\rangle}{#1}%
\DeclarePairedDelimiterX\innerprod[2]{\bigl(}{\bigr)}{{#1},\hspace*{0.2ex}{#2}}%
\DeclarePairedDelimiterX\CG[3]{\langle}{\rangle}{#1\,;\,#2\,\delimsize\vert\,#3}%
\DeclarePairedDelimiterX\scalprod[2]{\langle}{\rangle}{\,\vec#1\,,\,#2\,}%

\makeatother

\allowdisplaybreaks

\begin{document}
\preprint{\tt LPT-Orsay-18-41}
\preprint{\tt PCCF RI 1801}

\vspace*{22mm}

\title{Some hadronic parameters of charmonia in $\boldsymbol{N_{\text{f}}=2}$ lattice QCD}

\author{Gabriela~Bailas}
\affiliation{Laboratoire de Physique de Clermont\footnote[3]{Unit\'e Mixte de Recherche 6533 CNRS/IN2P3 -- Universit\'e Blaise Pasclal}, 4 Avenue Blaise Pascal, TSA 60026, 63171 Aubi\`ere Cedex, France}
\author{Beno\^it~Blossier}
\affiliation{Laboratoire de Physique Th\'eorique\footnote[1]{Unit\'e Mixte de Recherche 8627 du Centre National de la Recherche Scientifique}, CNRS, Univ. Paris-Sud et Universit\'e Paris-Saclay,  B\^atiment 210,  91405 Orsay Cedex, France}
\author{Vincent~Mor\'enas}
\affiliation{Laboratoire de Physique de Clermont\footnote[3]{Unit\'e Mixte de Recherche 6533 CNRS/IN2P3 -- Universit\'e Blaise Pasclal}, 4 Avenue Blaise Pascal, TSA 60026, 63171 Aubi\`ere Cedex, France}

\begin{abstract}
The phenomenology of leptonic decays of quarkonia holds many interesting features: for instance, it can establish constraints on scenarios beyond
the Standard Model with the Higgs sector enriched by a light CP-odd state. In the following paper, we report on a two-flavor lattice QCD study of the $\eta_c$ and $J/\psi$ decay constants, $f_{\eta_c}$ and $f_{J\psi}$. We also examine some properties of the first radial excitation $\eta_c(2S)$ and $\psi(2S)$.

\end{abstract}

\maketitle

\section{\label{Introduction}Introduction}

The discovery at LHC of the Higgs boson with a mass of 125.09(24) GeV \cite{AadZHL} has been a major milestone in the history of Standard Model (SM) tests: the spontaneous breaking of electroweak symmetry generates masses of charged leptons, quarks and weak bosons. A well-known issue with the SM Higgs is that the quartic term in the Higgs Lagrangian induces for the Higgs mass $m_H$ a quadratic divergence with the hard scale of the theory: it is related to the so-called hierarchy problem. Several scenarios beyond the SM are proposed to fix that
 theoretical caveat. Minimal extensions of the Higgs sector contain two complex scalar isodoublets $\Phi_{1,2}$ which, after the spontaneous breaking of the electroweak symmetry, lead to 2 charged particles $H^\pm$, 2 CP-even particles $h$ (SM-like Higgs) and $H$, and 1 CP-odd particle $A$. In that class of scenarios, quarks are coupled to the CP-odd Higgs through a pseudoscalar current. Those extensions of the Higgs sector have interesting phenomenological implications, especially as far as pseudoscalar quarkonia are concerned. For example, their leptonic decay is highly suppressed in the SM because it occurs \emph{via} quantum loops but it can be reinforced by the new tree-level contribution involving the CP-odd Higgs boson, in particular in the region of parameter space where the new boson is light ($10\, {\rm GeV} \lesssim m_A \lesssim 100\, {\rm GeV}$) and where the ratio of vacuum expectation values $\tan \beta$ is small ($\tan \beta <10$) \cite{FullanaUQ, BecirevicCHD}. Any enhanced observation with respect to the SM expectation would be indeed a clear signal of
New Physics. Let us finally note that the hadronic inputs, which constrain the CP-odd Higgs coupling to heavy quarks through processes involving quarkonia, are the decay constants $f_{\eta_c}$ and $f_{\eta_b}$. 
\par
This paper reports an estimate of hadronic parameters in the charmonia sector using lattice QCD with $N_f=2$ dynamical quarks: namely, the pseudoscalar decay constant $f_{\eta_c}$, because of its phenomenological importance, but also the vector decay constant $f_{J/\psi}$ as well as the ratio of masses $m_{\eta_c(2S)}/m_{\eta_c}$ and $m_{\psi(2S)}/m_{J/\psi}$. The two latter quantities are very well measured by experiments and their estimation has helped us to understand how much our analysis method can address the systematic effects, on quarkonia physics, coming from the lattice ensembles we have considered. We present also our findings for the following ratios of decay constants: $f_{\eta_c(2S)}/f_{\eta_c}$ and $f_{\psi(2S)}/f_{J/\psi}$. 
\par
This work is an intermediary step before going to the bottom sector, the final target of our program, because it is more promising for the phenomenology of extended Higgs sectors. An extensive study of the spectoscopy of charmonia has been done in \cite{LiuZE, LangSBA} while only two lattice estimates of $f_{\eta_c}$ and $f_{J/\psi}$ are available so far at ${\rm N_f=2}$ \cite{BecirevicBSA} and ${\rm N_f}=2+1$ 
\cite{DaviesIP, DonaldGA}.

\section{Lattice computation}
 
\renewcommand{\arraystretch}{1.1}
\begin{table}[t]
\begin{center}
\begin{tabular}{lcc@{\hskip 02em}c@{\hskip 02em}c@{\hskip 01em}c@{\hskip 01em}c@{\hskip 01em}c@{\hskip 01em}c}
\hline
	\toprule
	id	&	$\quad\beta\quad$	&	$(L/a)^3\times (T/a)$ 		&	$\kappa_{\rm sea}$		&	$a~(\rm fm)$	&	$m_{\pi}~(\MeV)$	& $Lm_{\pi}$ 	& $\#$ cfgs&$\kappa_c$\\
\hline
	\midrule
	E5	&	5.3		&	$32^3\times64$	& 	$0.13625$	& 	0.065	  	& 	$440$	&4.7	& $200$&$0.12724$\\  
	F6	&			& 	$48^3\times96$	&	$0.13635$	& 			& 	$310$	&5	& $120$&$0.12713$\\    
	F7	&			& 	$48^3\times96$	&	$0.13638$	& 			& 	$270$	&4.3	& $200$&$0.12713$\\    
	G8	&			& 	$64^3\times128$	&	$0.13642$	& 			& 	$190$	&4.1	& $176$&$0.12710$\\    
\hline
	\midrule
	N6	&	$5.5$	&	$48^3\times96$	&	$0.13667$	& 	$0.048$  	& 	$340$	&4	& $192$&$0.13026$\\	
	O7	&		&	$64^3\times128$	&	$0.13671$	& 	 	& 	$270$	&4.2	& $160$&$0.13022$ \\ 
	\bottomrule
\hline
\end{tabular} 
\end{center}
\caption{Parameters of the simulations: bare coupling $\beta = 6/g_0^2$, lattice resolution, hopping parameter $\kappa$, lattice spacing $a$ in physical units, pion mass, number of gauge configurations and bare charm quark masses.}
\label{tabsim}
\end{table}

\subsection{Lattice set-up}

This study has been performed using a subset of the CLS ensembles. These ensembles were generated with $N_f=2$ nonperturbatively $\mathcal{O}(a)$-improved Wilson-Clover fermions~\cite{SheikholeslamiIJ, LuscherUG} and the plaquette gauge action~\cite{WilsonSK} for gluon fields, by using either the DD-HMC algorithm~\cite{LuscherQA, LuscherRX, LuscherES, Luscherweb} or the MP-HMC algorithm~\cite{MarinkovicEG}. We collect in Table~\ref{tabsim} our simulation parameters. Two lattice spacings $a_{\beta=5.5}=0.04831(38)$ fm and $a_{\beta=5.3}=0.06531(60)$ fm, resulting from a fit in the chiral sector~\cite{LottiniRFA}, are considered. We have taken simulations with pion masses in the range $[190\,, 440]~\MeV$. The charm quark mass has been tuned after a linear interpolation of $m^2_{D_s}$ in $1/\kappa_c$ at its physical value \cite{HeitgerOAA}, after the fixing of the strange quark mass \cite{Fritzsch:2012wq}. The statistical error on raw data is estimated from the jackknife procedure: two successive measurements are sufficiently separated in trajectories along the Monte-Carlo history to neglect autocorrelation effects. Moreover, statistical errors on quantities extrapolated at the physical point are computed as follows. Inspired by the bootstrap prescription, we perform a large set of $N_{\rm event}$ fits of vectors of data whose dimension is the number of CLS ensembles used in our analysis ({\em i.e.}$\ n=6$) and where each component $i$ of those vectors is filled with an element randomly chosen among the $N_{\rm bins}(i)$ binned data per ensemble. The variance over the distribution of those $N_{\rm event}$  fit results, obtained with such ``random'' inputs, is then an estimator of the final statistical error. Finally, we have computed quark propagators through two-point correlation functions using stochastic sources which are different from zero in a single timeslice that changes randomly for each measurement. We have also applied spin dilution and the one-end trick to reduce the stochastic noise~\cite{FosterWU, McNeileBZ}.
\subsubsection{GEVP discussion}
\begin{figure}[htb]
	\begin{minipage}[c]{0.45\linewidth}
	\centering 
	\includegraphics*[width=0.9\linewidth]{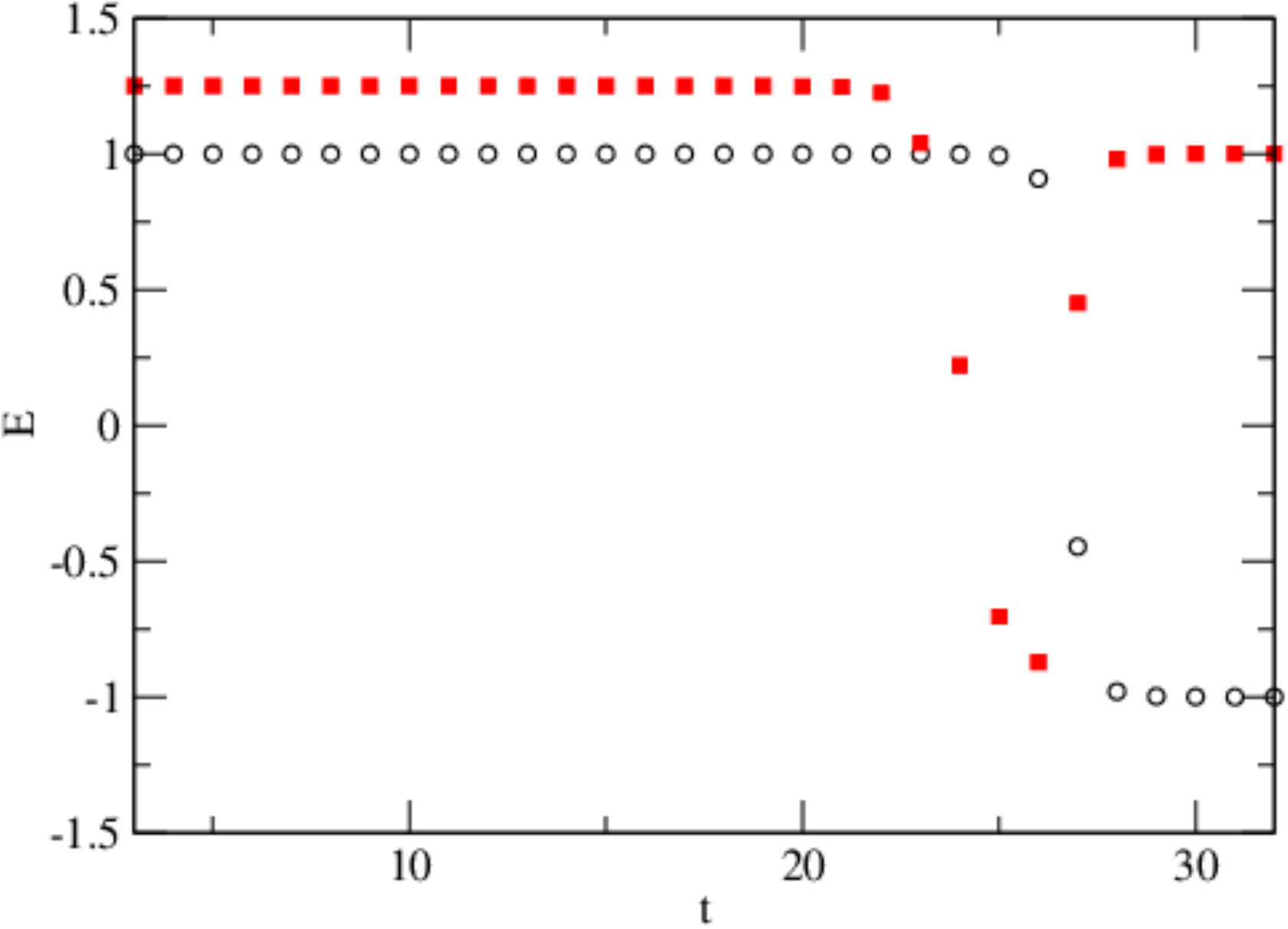}
	\end{minipage}
	\begin{minipage}[c]{0.45\linewidth}
	\centering 
	\includegraphics*[width=0.9\linewidth]{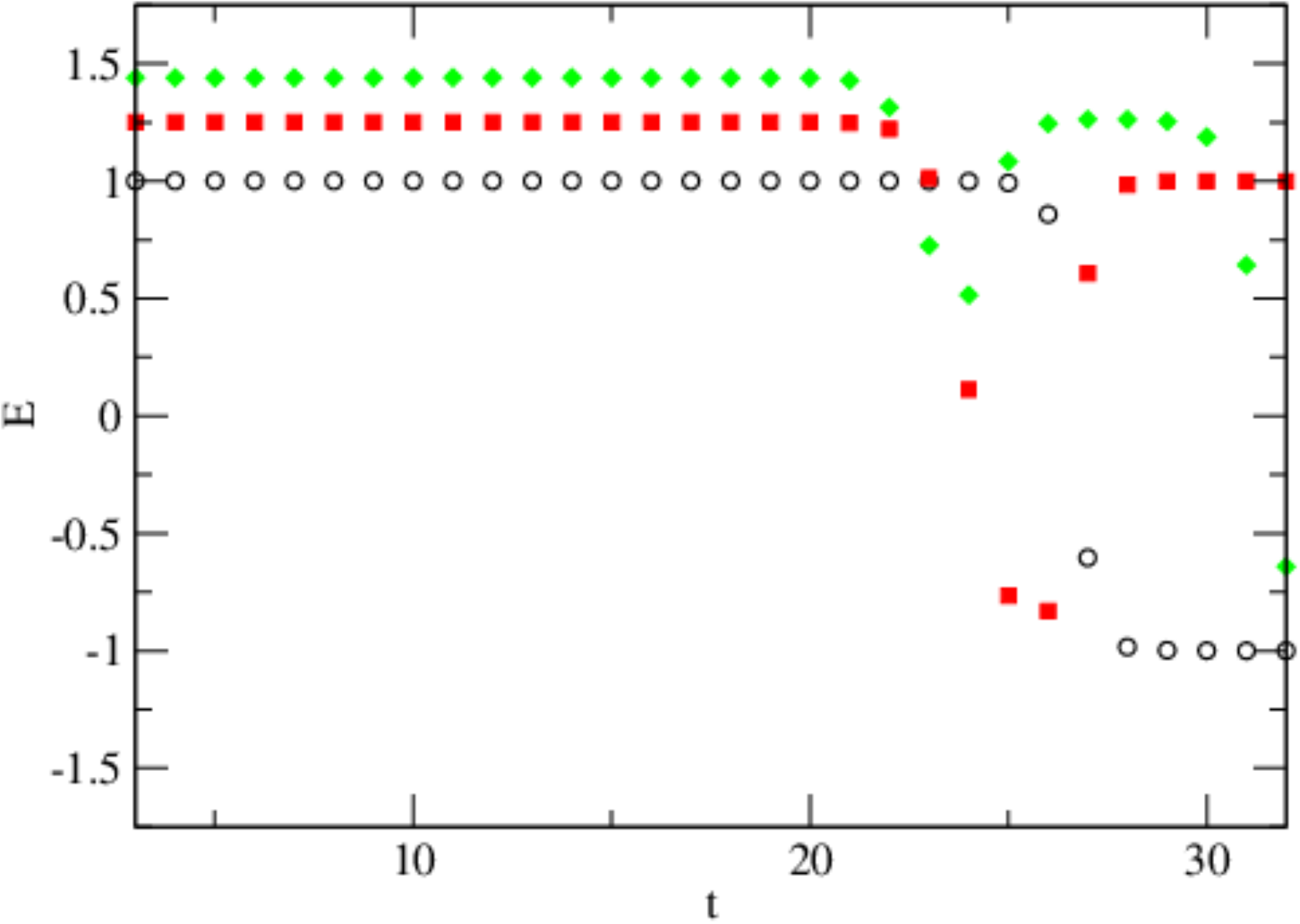}
	\end{minipage}
	\caption{Effective masses obtained from the $2\times 2$ subsystem (left panel) and the $3 \times 3$ subsystem (right panel) of our toy model, with $T=64$ and $t_0=3$.}
\label{fig:toymodel}
\end{figure}

The two-point correlation functions under investigation read
\begin{equation*}
C_{\Gamma \Gamma'}(t)=\frac{1}{V} \sum_{\vec{x},\vec{y}} \braet{\bigl[\bar{c}\,\Gamma c\bigr](\vec{y},t)}{\bigl[\bar{c}\,\gamma_0\,\Gamma' \gamma_0\,c\bigr](\vec{x},0)}
\end{equation*} 
where $V$ is the spatial volume of the lattice, $\langle ... \rangle$ is the expectation value over gauge configurations, and the interpolating fields $\bar{c} \Gamma c$ can be non-local.  As a preparatory work, different possibilities were explored to find the best basis of operators, combining levels of Gaussian smearing, interpolating fields with a covariant derivative $\bar{c} \Gamma (\vec{\gamma}\!\cdot\!\vec{\nabla}) c$ and operators that are odd under time parity. Solving the Generalized Eigenvalue Problem (GEVP)~\cite{MichaelNE, LuscherCK} is a key point in this analysis. Looking at the literature we have noticed that people tried to mix together the operators $\bar{c} \Gamma c$ and $\bar{c} \gamma_0 \Gamma c$ in a unique GEVP system \cite{LiuZE, BecirevicBSA}. In our point of view, this approach raises some questions: let us take the example of the interpolating fields $\{P=\bar{c} \gamma_5 c\,;\, A_0=\bar{c} \gamma_0 \gamma_5 c\}$. The asymptotic behaviours of the 2-pt correlation functions defined with these interpolating fields read 
\[
\left\{
\begin{aligned}
\braet{P(t)}{P(0)}\,;\,\braet{A_0(t)}{A_0(0)}\quad\xrightarrow{\ t \to \infty\ }\quad{\rm cosh}[m_P(T/2-t)]\\
\braet{P(t)}{A_0(0)}\,;\,\braet{A_0(t)}{P(0)}\quad\xrightarrow{\ t \to \infty\ }\quad{\rm sinh}[m_P(T/2-t)]
\end{aligned}
\right.  
\]
The matrix of $2\times 2$ correlators of the GEVP is then 
\[
C(t)=\begin{bmatrix}
\braet{P(t)}{P(0)}&\braet{A_0(t)}{P(0)}\\[2mm]
\braet{P(t)}{A_0(0)}&\braet{A_0(t)}{A_0(0)}
\end{bmatrix}
\quad {\rm GEVP}: 
C(t)\,v_n(t,t_0)\ =\ \lambda_n(t,t_0)\,C(t_0)\,v_n(t,t_0).
\]
In the general case, the spectral decomposition of $C_{ij}(t)$ is 
\beq\nonumber
C_{ij}(t)=\sum_n Z^i_n Z^{*j}_n [D_{ij}\rho^{(1)}_n(t) + (1-D_{ij}) \rho^{(2)}_n(t)]\quad\text{where}\quad
D_{ij}=0\ \text{or}\ 1 
\eeq
and with
\[
  \rho^{(1),(2)}(t) \sim e^{-m_Pt}\,{\rm cosh}[m_P(T/2-t)]\quad\text{or}\quad e^{-m_Pt}\,{\rm \sinh}[m_P(T/2-t)]
\]
The dual vector $u_n$ to $Z's$ is defined by
\beq\nonumber
\sum_j Z^{*j}_m u^j_n = \delta _{mn}
\eeq
Inserted in the GEVP, it gives
\bea\nonumber
\sum_j C_{ij}(t)u^j_n &=&\sum_{j,m} Z^i_m Z^{*j}_m u^j_n \bigl[D_{ij}\rho^{(1)}_m(t) + (1-D_{ij}) \rho^{(2)}_m(t) \bigr]\\
\nonumber
&=&\sum_m \rho^{(2)}_m(t) Z^i_m \sum_j Z^{*j}_m u^j_n + \sum_m \bigl(\rho^{(1)}_m(t)-\rho^{(2)}_m(t)\bigr)Z^i_m \sum_j D_{ij} Z^{*j}_m u^j_n\\
\nonumber
&=&\rho^{(2)}_n(t) Z^i_n + \sum_m \bigl(\rho^{(1)}_m(t)-\rho^{(2)}_m(t)\bigr)\,Z^i_m \sum_j D_{ij} Z^{*j}_m u^j_n
\eea
If $D_{ij}$ is independent of $i,j$, we can write 
\beq\nonumber
C(t)u_n = \rho(t) Z_n\quad\text{and}\quad \lambda_n(t,t_0)=\frac{\rho_n(t)}{\rho_n(t_0)}
\eeq
However, in the case of mixing T-odd and T-even operators in a single GEVP, the $D$'s \emph{do depend} on $i$ and $j$: the previous formula for $\lambda_n(t,t_0)$ is then not correct.
Hence, approximating every correlator by sums of exponentials forward in time, together with the assumption that the $D_{ij}$ are independent of $i$ and $j$, may face caveats. A toy model with 3 states in the spectrum helps to understand this issue:
\begin{center}
\begin{tabular}{|c|}
\hline
spectrum\\
\hline
1.0\\
1.25\\
1.44\\
\hline
\end{tabular}$\qquad\qquad$
\begin{tabular}{c}
Matrix of couplings\\
$\left[ \begin{array}{ccc}
0.6&0.25&0.08\\
0.61&0.27&0.08\\
0.58&0.24&0.08\\
\end{array}
\right]$
\end{tabular}$\qquad\qquad$
\begin{tabular}{c}
time behaviour of $C_{ij}$\\
$\left[ \begin{array}{ccc}
\cosh&\sinh&\cosh\\
\sinh&\cosh&\sinh\\
\cosh&\sinh&\cosh\\
\end{array}
\right]$
\end{tabular}
\end{center}
The effective energy $E^{\text{eff}}$ obtained by solving the GEVP writes
\beq\nonumber
aE^{\text{eff}}_{n}=\ln\left(\frac{\lambda_n(t,t_0)}{\lambda_n(t+a,t_0)}\right)
\eeq
In our numerical application, we have chosen $T=64$, $t_0=3$ and compared $2\times 2$ and $3 \times 3$ subsystems: results can be seen in Fig.\ref{fig:toymodel}.
Our observation is that until $t=T/4$, neglecting the time-backward contribution in the correlation function has no effect. Based on this study, we can affirm that the method is safe for the ground state and the first excitation. In contrast, one may wonder what would happen with a dense spectrum
 when the energy of the third or a higher excited state is extracted: in that case there will be a competition between the contamination from higher states, which are not properly isolated by the finite GEVP system, and the omission of the backward in time contribution to the generalised eigenvalue under study, even at times $t < 1$ fm. 
\subsubsection{Interpolating field basis}
Building a basis of operators with 
\[
  \bigl\{\bar{c}\,\Gamma c\ ;\ \bar{c}\, \Gamma (\vec{\gamma}\! \cdot\! \vec{\nabla}) c\bigr\}
  \qquad\text{where}\qquad
  \nabla_\mu \psi(x) =\frac{U_\mu(x)\psi(x+a\hat{\mu}) - U^\dag_\mu(x-a\hat{\mu})\psi(x-a\hat{\mu})}{2a}
\]
can have advantages and this was already explored~\cite{DudekWM, MohlerNA}. But there are sometimes bad surprises\ldots a good example being the pseudoscalar-pseudoscalar correlator defined by
\[
  C(t)\ =\ \braet{\bigl[\bar{c}\,\gamma^5 (\vec{\gamma}\!\cdot\!\vec{\nabla}) c\bigr](t)}{\bigl[\bar{c}\,\gamma^5 (\vec{\gamma}\!\cdot\!\vec{\nabla}) c\bigr](0)} 
\]
whose behaviour will be anticipated using the quark model method.\par\noindent
In the formalism of quark models, the $c$ quark field reads
\beq\nonumber
c(\vec{p}_c)\ =\ \begin{pmatrix}
  c_1\\[2mm]
  c_2=\dfrac{\vec{\sigma} \cdot \vec{p}_c}{2m_c}\,c_1
\end{pmatrix}
\eeq
being developped into a large and a small component. Notice that the $c$ quark field is a spinor of type $u$ while the $\bar c$ antiquark field is of type $v$, so at this stage we need to express $\bar v$ in terms of $u$. Defining the charge conjugation operator by ${\mathscr C}=-i\gamma^0 \gamma^2$ and using the Dirac representation for the $\gamma$ matrices, we can write
\[
v = {\mathscr C}(\bar{u})^T\qquad\Longrightarrow\qquad\bar v = (u)^T\gamma^0{\mathscr C}^\dagger\gamma^0
\]
The $\bar c$ antiquark field is then
\[
\bar c(\vec{p}_{\bar c})\ =\ -i\begin{pmatrix}
c^T_1\,{\dfrac{\vec\sigma^T\cdot\vec p_{\bar c}}{2m_c}}\,\sigma_2\ &\ c^T_1\,\sigma_2
\end{pmatrix}
\]
Assuming the charmonium at rest, we have 
\begin{equation}\nonumber
\vec{p}_c=-\vec{p}_{\bar{c}}\qquad\text{and}\qquad (\vec{\sigma}\cdot \vec{p}_{\bar{c}})^T \sigma_2 = \sigma_2 (\vec{\sigma}\cdot \vec{p}_c)
\end{equation}
Then ordinary Pauli matrices algebra leads to
\begin{align*}
  \bar{c}\,\gamma^5 (\vec{\gamma}\!\cdot\!\vec{\nabla}) c\ &=\ \sum\limits_{n=1}^3
-i\begin{pmatrix}c^T_1\,{\dfrac{\vec\sigma^T\cdot\vec{p}_{\bar{c}}}{2m_c}}\,\sigma_2\ &\ c^T_1\,\sigma_2\end{pmatrix}
\begin{pmatrix}-\sigma_n&0\\ 0&\sigma_n\end{pmatrix} D_n\begin{pmatrix}c_1\\[2mm] \dfrac{\vec\sigma\cdot\vec{p}_c}{2m_c}\,c_1 \end{pmatrix}\\
  &= \sum\limits_{n=1}^3 c_1^T\left[{{-\dfrac{\vec\sigma^T\cdot\vec{p}_{\bar{c}}}{2m_c}}\,\sigma_2\,\sigma_n + \sigma_2\,\sigma_n\,{\dfrac{\vec\sigma\cdot\vec{p}_c}{2m_c}}}\right]{p_c}_n\,c_1\qquad\text{assuming}\ D_n\sim i\,{p_c}_n\\
&=\sum\limits_{n=1}^3 c^T_{1}\sigma_2\, \dfrac{-(\vec{\sigma}\cdot \vec{p}_c )\,\sigma_n + \sigma_n\, (\vec{\sigma}\cdot \vec{p}_c)}{2m_c}\,{p_c}_n\,c_1\ 
=\ \sum\limits_{n,m=1}^3 c^T_{1}\sigma_2\, \dfrac{\sigma_n\,\sigma_m - \sigma_m\,\sigma_n}{2m_c}\,{p_c}_n\,{p_c}_m\,c_1\\
&=\sum\limits_{n,m,\ell=1}^3 i\,c^T_{1}\sigma_2\, \dfrac{\epsilon_{nm\ell}\,\sigma_\ell}{m_c}\,{p_c}_n\,{p_c}_m\,c_1\ =\ 0
\end{align*}
The previous approach is very naive with respect to quantum field theory, in particular the approximation $D_i c_1 \sim p_i c_1$, but as a conclusion one can see that
interpolating fields $\bar{c}\,\gamma^5 (\vec{\gamma}\!\cdot\!\vec{\nabla}) c$ potentially give very noisy correlators. And, indeed, we have found
a numerical cancellation between the ``diagonal'' contribution 
\[
A(t)=\sum_i \braet{\bigl[\bar{c}\,\gamma^5 \gamma_i \nabla_i\, c\bigr](t)}{\bigl[\bar{c}\, \gamma^5 \gamma_i \nabla_i\, c\bigr](0)}
\]
and the ``off-diagonal'' contribution 
\[
B(t)=\sum_{i\neq j} 
\braet{\bigl[\bar{c}\,\gamma^5 \gamma_i \nabla_i\, c\bigr](t)}{\bigl[\bar{c}\, \gamma^5 \gamma_j \nabla_j\, c\bigr](0)}
\]
resulting in a very noisy correlator $C(t)$ which is compatible with zero.
\subsubsection{Summary}
To summarise, we have considered four Gaussian smearing levels for the quark field $c$, including no smearing, to build $4\times 4$ matrix of correlators without any covariant derivative nor operator of the $\pi_2$ or $\rho_2$ kind \cite{DudekWM}, from which we have also extracted
 the ${\cal O}(a)$ improved hadronic quantities we have examined. Solving the GEVP for the pseudoscalar-pseudoscalar
and vector-vector matrices of correlators 
\begin{equation*}
C_{PP}(t)v^P_n(t,t_0)=\lambda^P_n(t,t_0) v^P_n(t,t_0)C_{PP}(t_0) \quad \text{and}\quad
C_{VV}(t)v^V_n(t,t_0)=\lambda^V_n(t,t_0) v^V_n(t,t_0)C_{VV}(t_0)
\end{equation*} 
we obtain the correlators that will have the largest overlap with the $n^{\rm th}$ excited state as follows:
\begin{align*}
\tilde{C}^n_{A_0 P}(t) &= \sum_i C_{A^L_0 P^{(i)}}(t) v^{P,i}_n(t,t_0)\\[2mm]
\tilde{C}^n_{P P}(t) &= \sum_i C_{P^L P^{(i)}}(t) v^{P,i}_n(t,t_0)\\[2mm]
\tilde{C}'^n_{P P}(t) &= \sum_{i,j} v^{P,i}_n(t,t_0) C_{P^{(i)} P^{(j)}}(t) v^{P,j}_n(t,t_0)\\[2mm]
\tilde{C}^n_{V V}(t) &= \frac{1}{3} \sum_{i,k} C_{V^L_k V^{(i)}_k}(t) v^{V,i}_n(t,t_0)\\[2mm]
\tilde{C}'^n_{VV}(t) &= \frac{1}{3} \sum_{i,j,k} v^{V,i}_1(t,t_0) C_{V^{(i)}_k V^{(j)}_k}(t) v^{V,j}_n(t,t_0)\\[2mm]
\tilde{C}^n_{TV}(t) &= \frac{1}{3} \sum_{i,k} C_{T^L_{k0} V^{(i)}_k}(t) v^{V,i}_n(t,t_0)\\[2mm]
\tilde{C}^n_{\delta PP}(t) &= \frac{\tilde{C}^n_{PP}(t+1)-\tilde{C}^n_{PP}(t-1)}{2a}\\[2mm]
\tilde{C}^n_{\delta TV}(t) &= \frac{\tilde{C}^n_{TV}(t+1) - \tilde{C}^n_{TV}(t-1)}{2a}
\end{align*}
and their symmetric counterpart with the exchange of operators at the source and at the sink. The quark bilinears are $P=\bar{c}\gamma_5 c$, $A_0=\bar{c}\gamma_0\gamma_5 c$, $V_k=\bar{c}\gamma_k c$ and $T_{k0}=\bar{c}\gamma_k\gamma_0 c$. Moreover, in those expressions,
 the label $L$ refers to a local interpolating field while sums over $i$ and $j$ run over the four Gaussian smearing levels.\par\noindent
 The projected correlators have the following asymptotic behaviour
\begin{align*}
\tilde{C}'^n_{P P}(t)&\xrightarrow[\ t/a \gg 1\ ]{} \frac{{\cal Z}'_{{PP}_n}}{am_{P_n}} e^{-m_{P_n} T/2} \cosh [m_{P_n} (T/2 -t)]\\[2mm]
\tilde{C}^n_{A_0 P}(t)&\xrightarrow[\ t/a \gg 1\ ]{} -\frac{{\cal Z}_{{AP}_n}}{am_{P_n}}e^{-m_{P_n} T/2} \sinh [m_{P_n} (T/2-t )]\\[2mm]
\tilde{C}^n_{P P}(t)&\xrightarrow[\ t/a \gg 1\ ]{}\frac{{\cal Z}_{{PP}_n}}{am_{P_n}} e^{-m_{P_n} T/2} \cosh [m_{P_n} (T/2 -t)]\\[2mm]
\tilde{C}'^n_{VV}(t)&\xrightarrow[\ t/a \gg 1\ ]{} \frac{{\cal Z}'_{{VV}_n}}{am_{V_n}} e^{-m_{V_n} T/2} \cosh [m_{V_n} (T/2 -t)]\\[2mm]
\tilde{C}^n_{VV}(t)&\xrightarrow[\ t/a \gg 1\ ]{} \frac{{\cal Z}_{{VV}_n}}{am_{V_n}}e^{-m_{V_n} T/2} \cosh [m_{V_n} (T/2-t )]\\[2mm]
\tilde{C}^n_{T V}(t)&\xrightarrow[\ t/a \gg 1\ ]{} \frac{{\cal Z}_{{TV}_n}}{am_{V_n}} e^{-m_{V_n} T/2} \sinh [m_{V_n} (T/2 -t)]\\[2mm]
\tilde{C}^n_{\delta P P}(t)&\xrightarrow[\ t/a \gg 1\ ]{} -1/a\, \sinh (am_{P_n}) \frac{{\cal Z}_{{PP}_n}}{am_{P_n}} e^{-m_{P_n} T/2} \sinh [m_{P_n} (T/2 -t)]\\[2mm]
\tilde{C}^n_{\delta T V}(t)&\xrightarrow[\ t/a \gg 1\ ]{} -1/a\,\sinh (am_{V_n}) \frac{{\cal Z}_{{TV}_n}}{am_{V_n}} e^{-m_{V_n} T/2} 
\cosh [m_{V_n} (T/2 -t)]
\end{align*}
\subsubsection{Decay constant extraction}\noindent
Considering the ${\cal O}(a)$ improved operators 
\begin{equation*}
A^I_0=(1+b_A Z a m_{c}^{AWI})\left(A_0 + a c_A \frac{\partial_0+\partial^*_0}{2} P\right)
\quad\text{and}\quad 
V^I_k=(1+b_V Z a m_{c}^{AWI})\left(V_k + ac_V \frac{\partial_\nu + \partial^*_\nu}{2}T_{k\nu}\right)
\end{equation*}
where the lattice derivatives are defined by
\begin{equation*}
\partial_\nu F(x) = \frac{F(x+a\hat{\nu})-F(x)}{a}\qquad\text{as well as}\qquad \partial^*_\nu F(x)=\frac{F(x)-F(x-a\hat{\nu})}{a},
\end{equation*}
the $f_{\eta_c}$ decay constants is extracted in the following way:
\begin{equation*}
\left\{
  \begin{aligned}
    \quad&\brakket{0}{A^R_0}{\eta_c(\vec{p}=0)} = -f_{\eta_c} m_{\eta_c} = -Z_A(1 + b_A Z a m^{AWI}_c) m_{\eta_c} f^0_{P_c}(1 + f^1_{P_c}/f^0_{P_c})\\[2mm] 
\quad&af^0_{P_c}=\frac{1}{am_{\eta_c}} \frac{{\cal Z}_{{AP}_1}}{\sqrt{{\cal Z}'_{{PP}_1}}}\qquad;\qquad
af^1_{P_c} =  \frac{1}{am_{\eta_c}} c_A \sinh (a m_{\eta_c}) \frac{{\cal Z}_{{PP}_1}}{\sqrt{{\cal Z}'_{{PP}_1}}}
  \end{aligned}
\right.
\end{equation*}
while the $f_{J/\psi}$ decay constant is obtained with:
\begin{equation*}
  \left\{
    \begin{aligned}
      \quad&\brakket{0}{V^R_i}{J/\psi(\epsilon, \vec{p}=0)} = \epsilon_i f_{J/\psi} m_{J/\psi} = \epsilon_i Z_V(1 + b_V Z a m^{AWI}_c) m_{J/\psi} f^0_{V_c}(1 + f^1_{V_c}/f^0_{V_c})\\[2mm] 
  \quad&af^0_{V_c}=\frac{1}{am_{J/\psi}} \frac{{\cal Z}_{{VV}_1}}{\sqrt{{\cal Z}'_{{VV}_1}}}\qquad;\qquad af^1_{V_c} = - \frac{1}{am_{J/\psi}}c_V \sinh (a m_{J/\psi}) \frac{{\cal Z}_{{TV}_1}}{\sqrt{{\cal Z}'_{{VV}_1}}}
    \end{aligned}
  \right.
\end{equation*}
The $R$ superscript denotes the renormalized improved operators. The renormalization constants $Z_A$ and $Z_V$ have been non perturbatively measured in~\cite{DellaMorteXB, DellaMorteRD}. 
We have also used non-perturbative results and perturbative formulae from~\cite{DellaMorteAQE, SintJX, GuagnelliJW} for the improvement coefficients $c_A$, $c_V$, $b_A$ and $b_V$, and the matching coefficient $Z$ between the quark mass $m_{c}$ defined through the axial Ward Identity 
\[
m_{c}^{AWI}=\frac{\frac{\partial_0 +\partial^*_0}{2} C_{A^L_0P^L}(t) +ac_A \partial_0 \partial^*_0 C_{P^LP^L}(t)}{2C_{P^LP^L}(t)}
\]
and its counterpart defined through the vector Ward Identity 
\[
am_{c}^{VWI}=\frac{1}{2} \left(\frac{1}{\kappa_c} - \frac{1}{\kappa_{cr}}\right)
\]
The decay constants of the considered radial excited states are given by
\begin{equation*}
  \left\{
    \begin{aligned}
      \quad&\brakket{0}{A^R_0}{\eta_c(2S)(\vec{p}=0)} = -f_{\eta_c(2S)} m_{\eta_c(2S)} = -Z_A(1 + b_A Z a m^{AWI}_c) m_{\eta_c(2S)} f^0_{P'_c}(1 + f^1_{P'_c}/f^0_{P'_c})\\[2mm] 
  \quad&af^0_{P'_c}=\frac{1}{am_{\eta_c(2S)}} \frac{{\cal Z}_{{AP}_2}}{\sqrt{{\cal Z}'_{{PP}_2}}}\qquad;\qquad af^1_{P'_c} =  \frac{1}{am_{\eta_c(2S)}} c_A \sinh (a m_{\eta_c(2S)}) \frac{{\cal Z}_{{PP}_2}}{\sqrt{{\cal Z}'_{{PP}_2}}}
    \end{aligned}
  \right.
\end{equation*}
and
\begin{equation*}
  \left\{
    \begin{aligned}
      \quad&\brakket{0}{V^R_i}{\psi(2S)(\epsilon, \vec{p}=0)} = \epsilon_i f_{\psi(2S)} m_{\psi(2S)} = \epsilon_i Z_V(1 + b_V Z a m^{AWI}_c) m_{\psi(2S)} f^0_{V'_c}(1 + f^1_{V'_c}/f^0_{V'_c})\\[2mm] 
  \quad&af^0_{V'_c}=\frac{1}{am_{\psi(2S)}} \frac{{\cal Z}_{{VV}_2}}{\sqrt{{\cal Z}'_{{VV}_2}}}\qquad;\qquad af^1_{V'_c} = - \frac{1}{am_{\psi(2S)}}c_V \sinh (a m_{\psi(2S)}) \frac{{\cal Z}_{{TV}_2}}{\sqrt{{\cal Z}'_{{VV}_2}}}
    \end{aligned}
  \right.
\end{equation*}

\subsection{Analysis}

Let us first consider the mass and the decay constant of the ground states. Since the fluctuations with time are large, we have decided to use generalized eigenvectors \emph{at fixed time} $t_{\text{fix}}$, $v^{P(V)}_
1 (t_{\rm fix}, t_0)$, in order to perform the corresponding projection. In practice we have chosen $t_{\text{fix}}/a = t_0/a + 1$ but we have checked that the results do not depend on $t_{\text{fix}}$. However, for the excited states, the formulae written above apply. Although the fluctuations are even larger than for the ground states, the correlators $\tilde{C}^2_{A_0P}$ and $\tilde{C}^2_{VV}$ are qualitatively well fitted by a single contribution, which is not true at all if the projection on $v^{P(V)}_2(t_{\rm fix},t_0)$ is used: effective masses extracted with the latter projection show a very big slope with time.
For the ground states, the time range $[t_{\rm min},t_{\rm max}]$ used to fit the projected correlators is set so that the statistical error on the effective mass $\delta m^{\rm stat}(t_{\rm min})$ is larger enough than the systematic error $\Delta m^{\rm sys}(t_{\rm min})\equiv \exp[-\Delta  t_{\rm min}]$ with $\Delta  = E_4 - E_1  \sim 2\,{\rm GeV}$. That guesstimate is based on our $4\times 4$ GEVP analysis, though we do not want to claim here that we really control the energy-level of the third excited state. Actually our criterion is rather $\delta m^{\rm stat}(t_{\rm min})> 4\Delta m^{\rm sys}(t_{\rm min})$ to be more conservative. On the other side, $t_{\rm max}$ is set after a qualitative inspection of the data which count for the plateau determination. Since $\Delta^{(P)} \sim \Delta^{(V)}$, fit intervals are identical for pseudoscalar and vector charmonia. 
For the first excited states, the time range has been set by looking at effective masses and where the plateaus start and end, including statistical uncertainty. Finally, $t_0=3a$ at $\beta=5.3$ and $t_0=5a$ at $\beta=5.5$: the landscape is unchanged by increasing $t_0$. We have collected in Table~\ref{tabresults} of the Appendix the raw data we have obtained in our analysis.\par\noindent
The extrapolation to the physical point of the quantities we have measured uses a linear expression in $m^2_\pi$ with inserted cut-off effects in $a^2$:
\begin{equation*}
X(a,m_\pi)=X_0 + X_1 m^2_\pi + X_2 (a/a_{\beta=5.3})^2
\end{equation*} 
We remind that $\kappa_c$ have been tuned at every $\kappa_{\rm sea}$ so that $m_{D_s}(\kappa_s,\kappa_c,\kappa_{\rm sea})=m^{\rm phys}_{D_s}$ {(it was necessary to tune the strange quark mass $\kappa_s$, though it is not a relevant quantity for the study discussed here)}.
\par
\begin{figure}[t] 
	\begin{minipage}[c]{0.45\linewidth}
	\centering 
	\includegraphics*[width=0.9\linewidth]{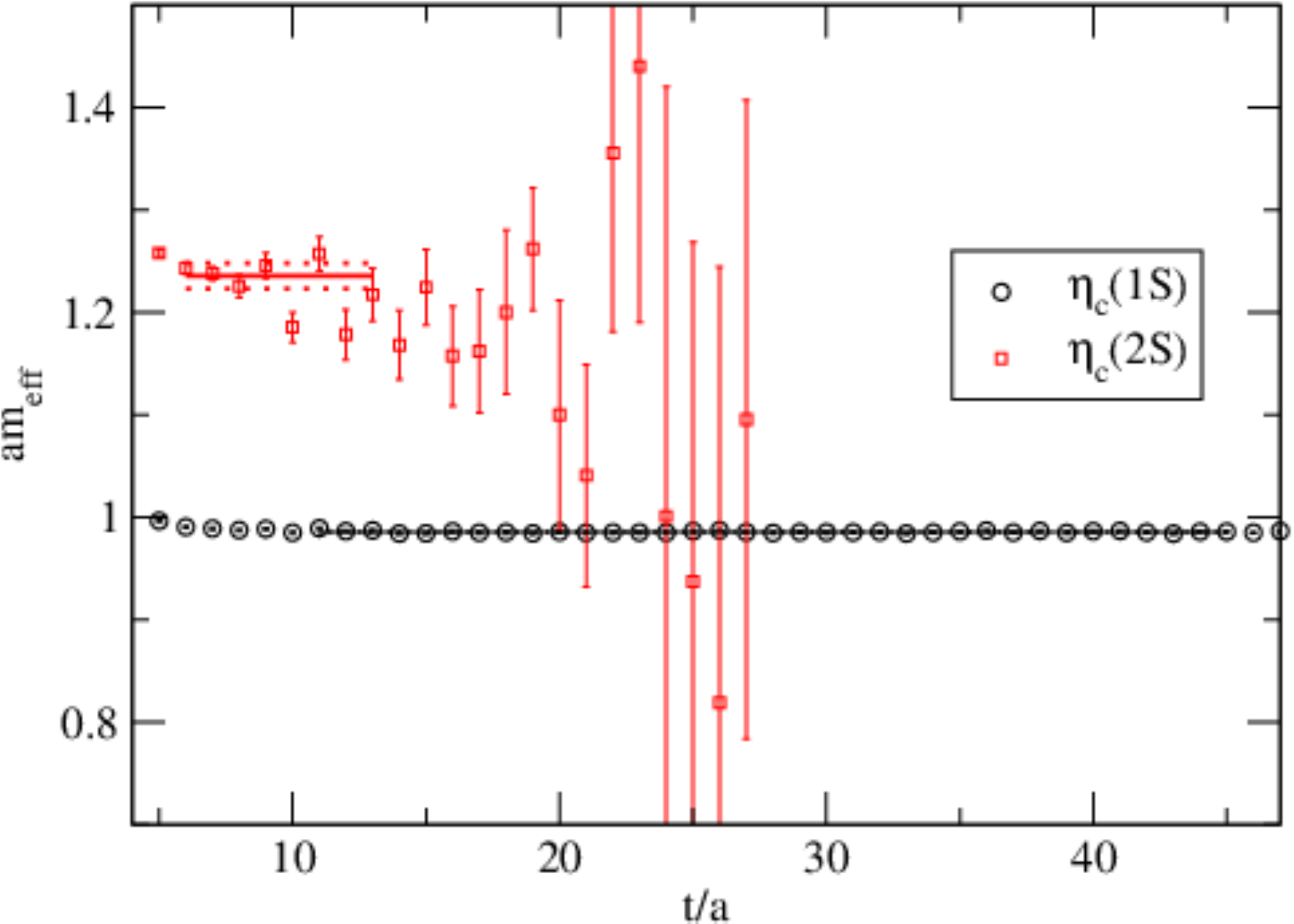}
	\end{minipage}
	\begin{minipage}[c]{0.45\linewidth}
	\centering 
	\includegraphics*[width=0.9\linewidth]{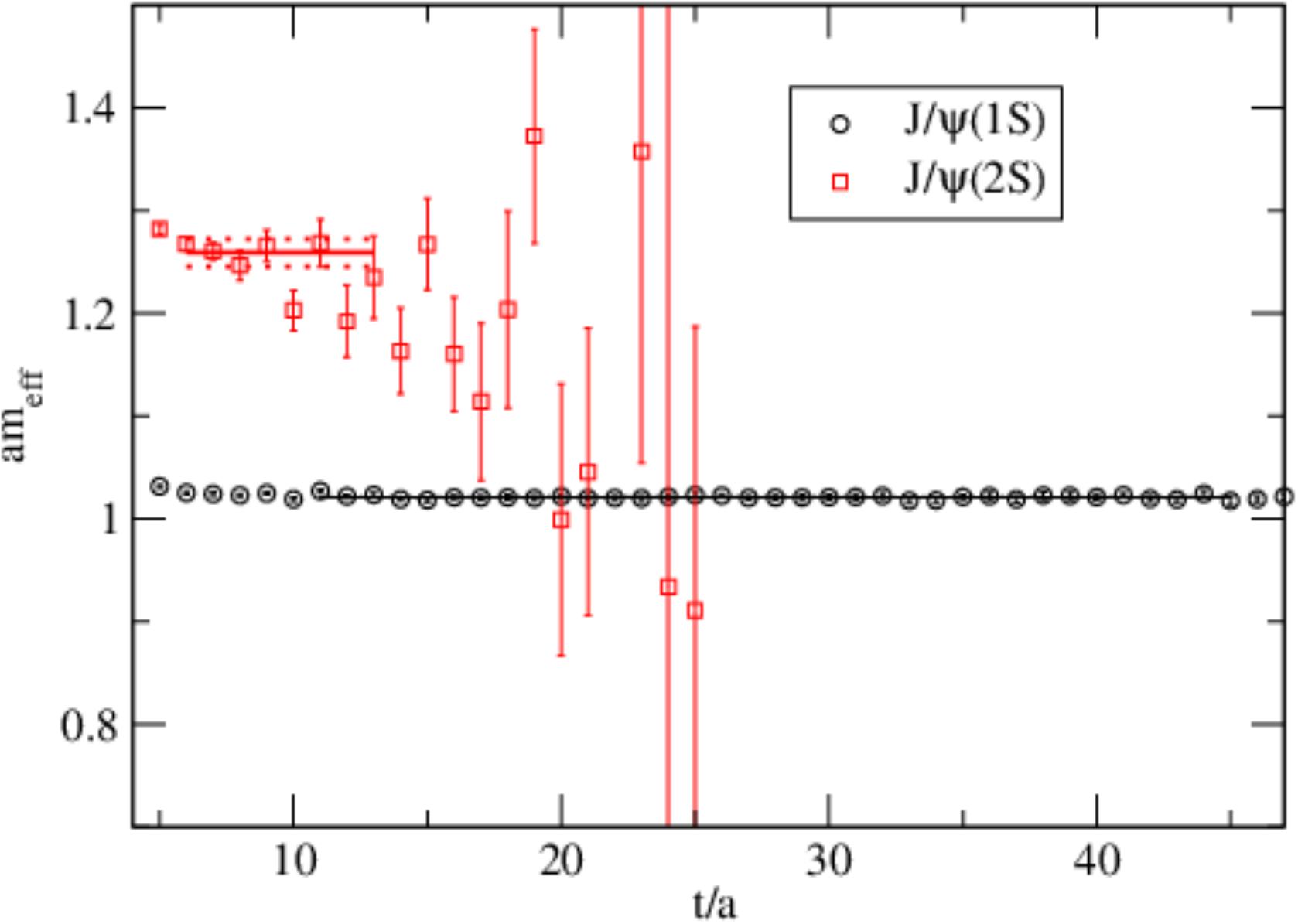}
	\end{minipage}
	\caption{Effective masses $am_{\eta_c}$ and $am_{\eta_c(2S)}$ (left panel), $am_{J/\psi}$ and $am_{\psi(2S)}$ (right panel) extracted from a $4\times 4$ GEVP for the lattice ensemble F7; we also plot the plateaus obtained in the chosen fit interval.}
\label{fig:masses}
\end{figure}

\noindent 
The Fig.\ref{fig:masses} shows the effective masses of the $\eta_c$, $\eta_c(2S)$, $J/\psi$ and $\psi(2S)$ mesons for the set F7 which are obtained by the following expressions
\[
\left\{
\begin{aligned}
am_{\eta_c}^{\rm eff}(t)&\ =\ {\rm argcosh} \left(\frac{\lambda^P_1(t+a,t_0)+\lambda^P_1(t-a,t_0)}{2\lambda^P_1(t,t_0)}\right)\\[2mm]
\quad  am_{J/\psi}^{\rm eff}(t)&\ =\ {\rm argcosh} \left(\frac{\lambda^V_1(t+a,t_0)+\lambda^V_1(t-a,t_0)}{2\lambda^V_1(t,t_0)}\right)
\end{aligned}
\right.
\]
and
\[
  \left\{
\begin{aligned}
am_{\eta_c(2S)}^{\rm eff}(t)&\ =\ {\rm argcosh} \left(\frac{\lambda^P_2(t+a,t_0)+\lambda^P_2(t-a,t_0)}{2\lambda^P_1(t,t_0)}\right)\\[2mm]
\quad  am_{\psi(2S)}^{\rm eff}(t)&\ =\ {\rm argcosh} \left(\frac{\lambda^V_2(t+a,t_0)+\lambda^V_2(t-a,t_0)}{2\lambda^V_1(t,t_0)}\right)
\end{aligned}
\right.
\]
The drawn lines correspond to our plateaus for the fitted masses. 
One can observe that our plateaus are large for the ground states but they are unfortunately of a much worse quality for the radial excitations. The latter data also show large fluctuations with time, for which we did not find any satisfying explanation. 
If $t_{\rm fix}$ were used in the correlators projection procedure,
the price to pay would be a strong contamination by other states in $\tilde{C}^2_{A_0P}$ and $\tilde{C}^2_{VV}$ other than the ones of interest.\par
\begin{figure}[t] 
	\begin{minipage}[c]{0.45\linewidth}
	\centering 
\includegraphics*[width=7cm, height=5cm]{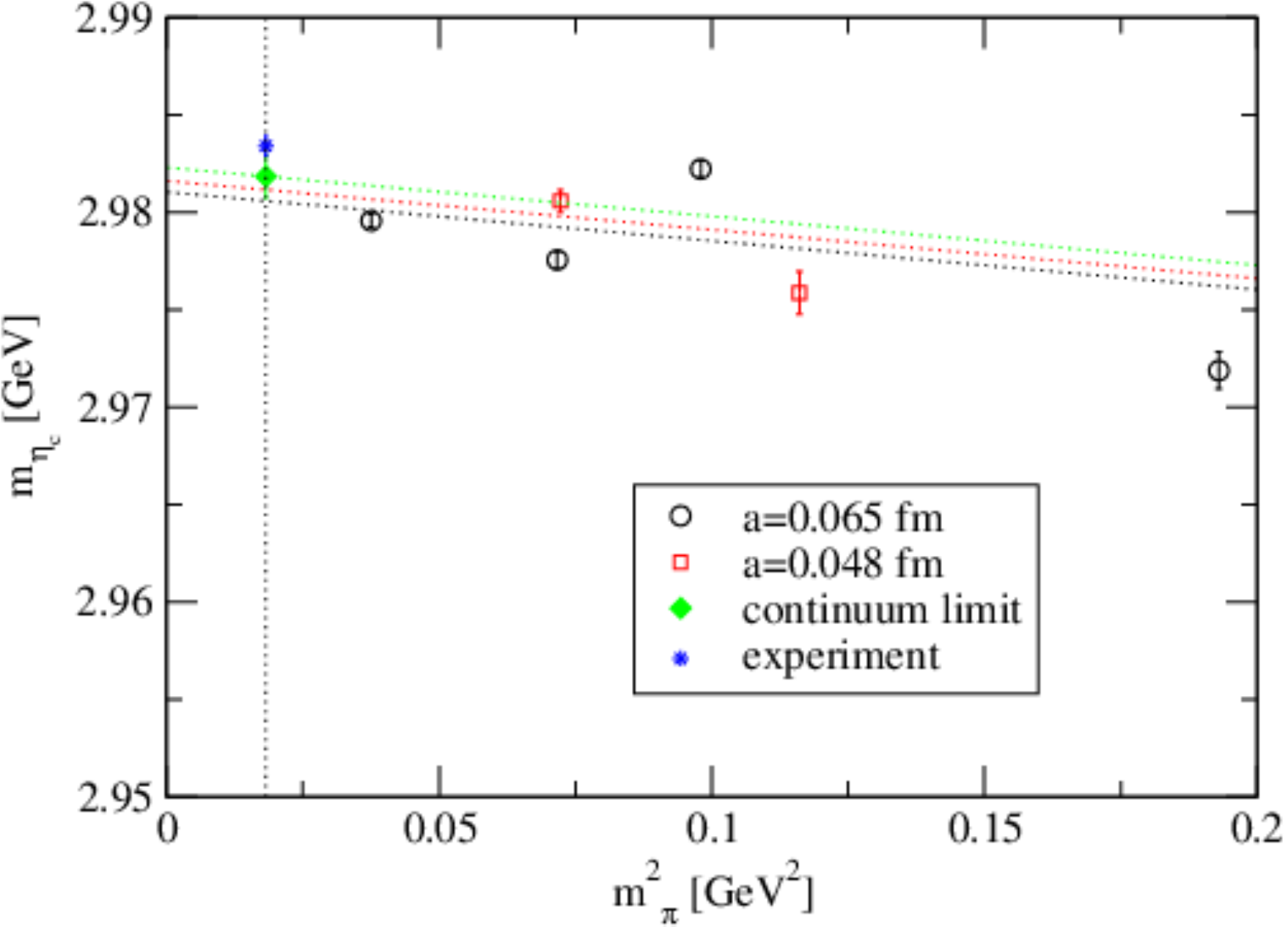}
	\end{minipage}
	\begin{minipage}[c]{0.45\linewidth}
	\centering 
\includegraphics*[width=7cm, height=5cm]{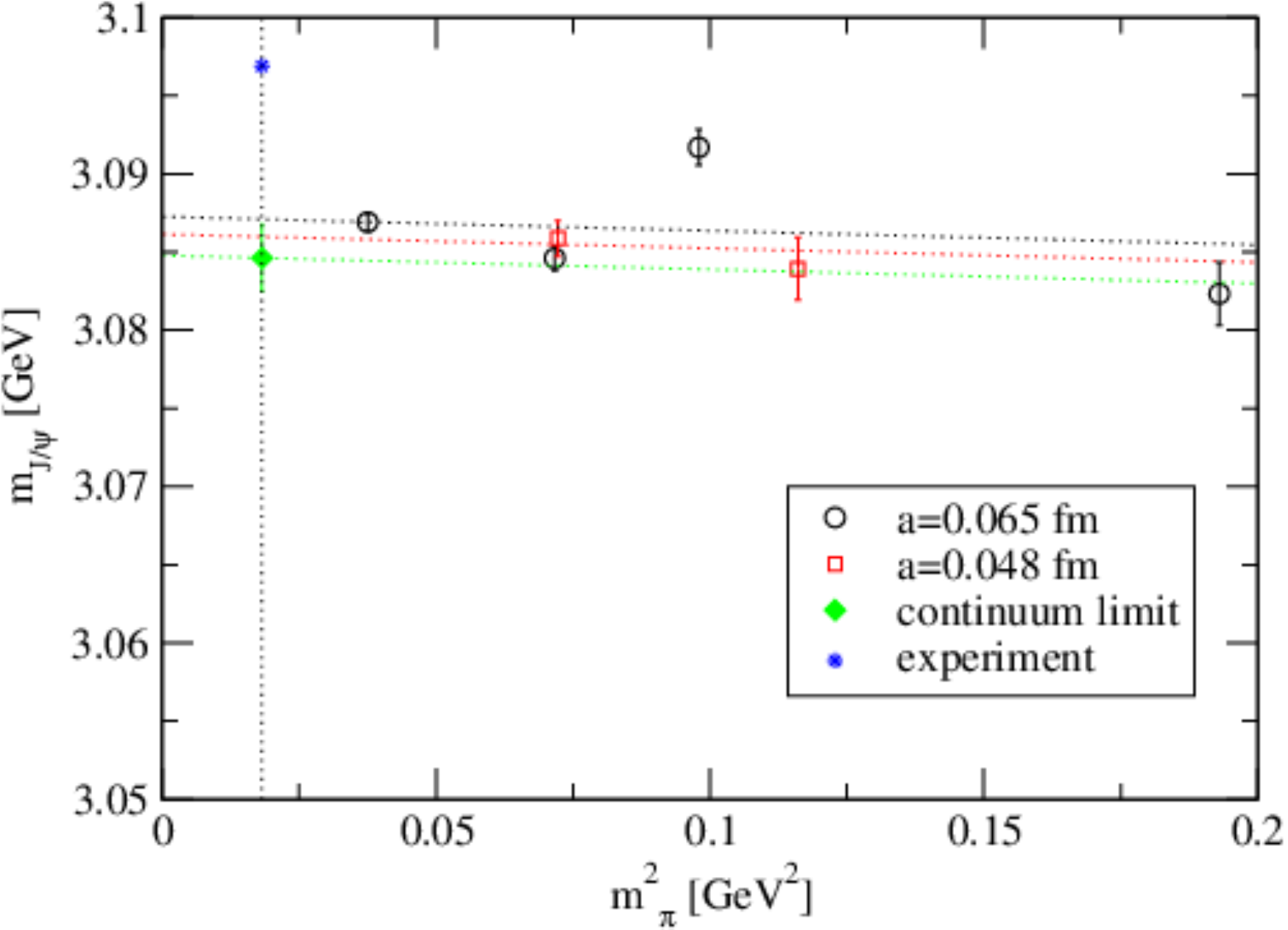}
	\end{minipage}
	\caption{Extrapolation at the physical point of $f_{\eta_c}$ (left panel) and $f_{J/\psi}$ (right panel) by linear expressions in $m^2_\pi$ and $a^2$.}
\label{fig:massphysical}
\end{figure}

\noindent
We show in Fig.\ref{fig:massphysical} the extrapolation to the physical point of $m_{\eta_c}$ and $m_{J/\psi}$. The dependence on $m^2_\pi$ and $a^2$ is mild, with cut-off effects almost negligible. However the contribution to the meson masses besides the mass term $2m_c$ is difficult to catch. At the physical point, $m_{\eta_c}$ and $m_{J/\psi}$ are compatible with the experimental values 2.983 GeV and 3.097 GeV: 
\begin{equation*}
m_{\eta_c}=2.982(1)(19)\,{\rm GeV}\quad\text{and}\quad m_{J/\psi}=3.085(2)(20)\, {\rm GeV}
\end{equation*}
where the first error is statistical and the second error accounts for
 the uncertainty on the lattice spacing. The latter clearly dominates and hides a possible mismatch between our extrapolated results at the physical point and experiment.\par
\begin{figure}[t] 
	\begin{minipage}[c]{0.45\linewidth}
	\centering 
\includegraphics*[width=7cm, height=5cm]{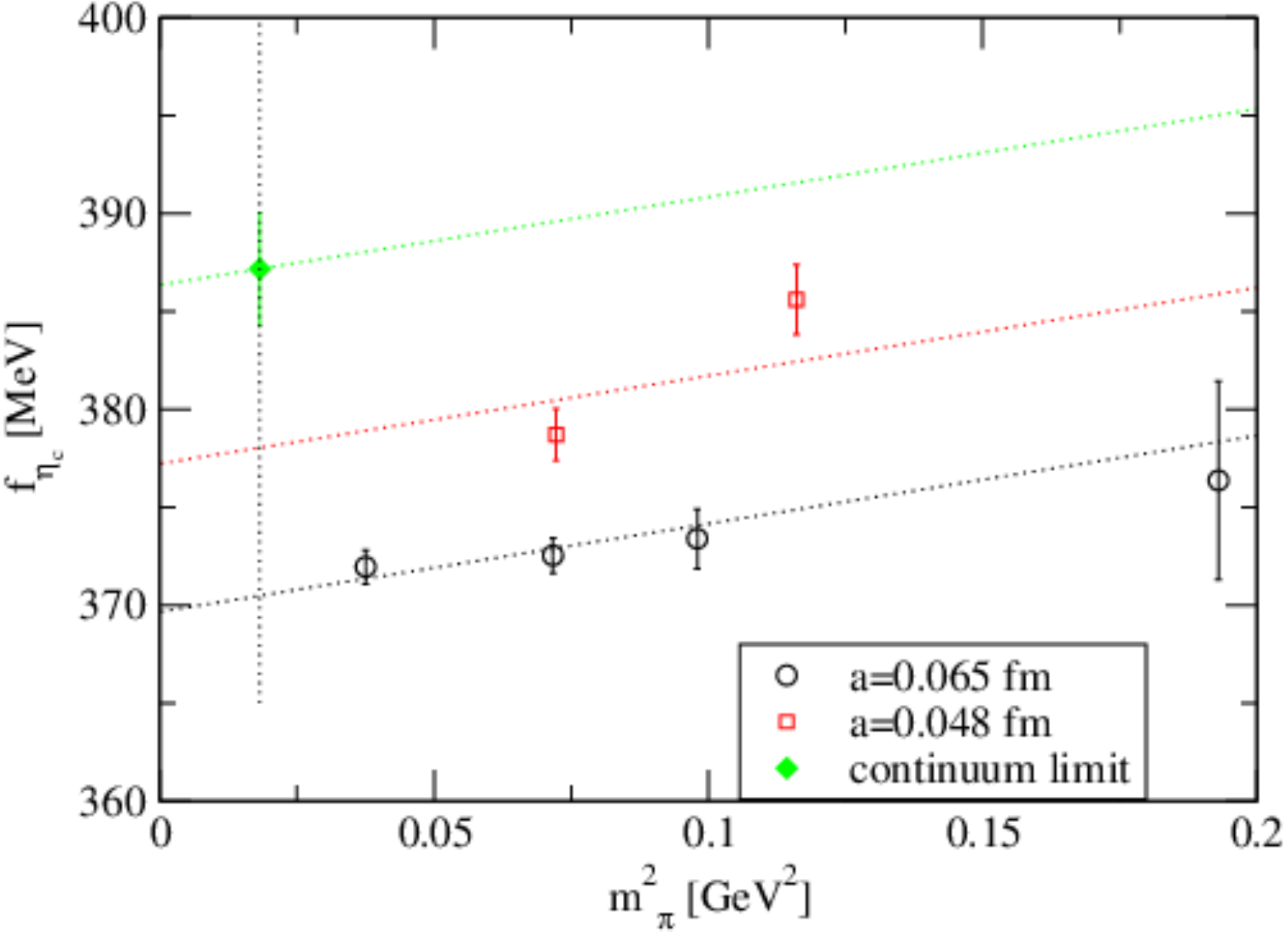}
	\end{minipage}
	\begin{minipage}[c]{0.45\linewidth}
	\centering 
\includegraphics*[width=7cm, height=5cm]{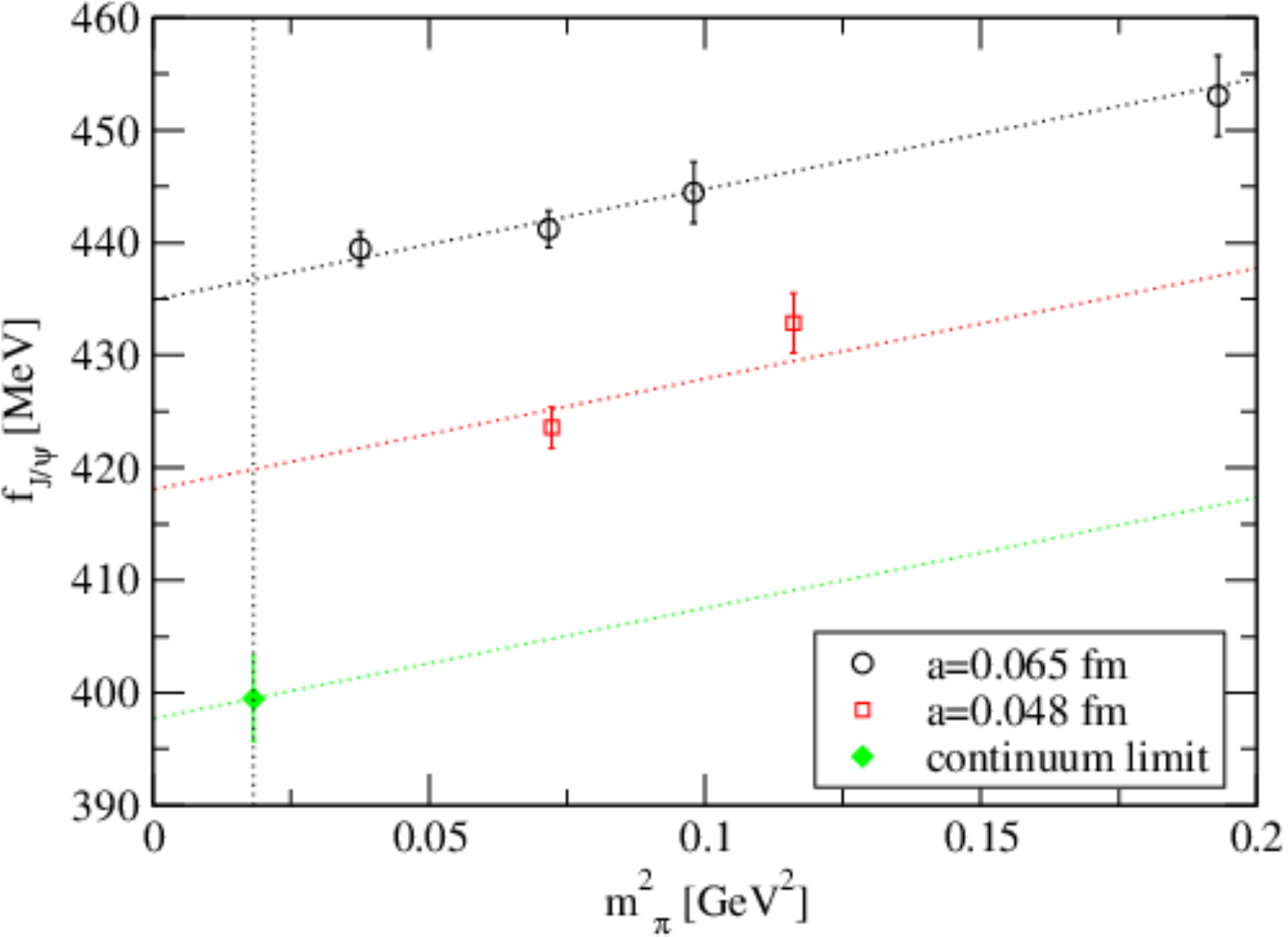}
	\end{minipage}
	\caption{Extrapolation at the physical point of $f_{\eta_c}$ (left panel) and $f_{J/\psi}$ (right panel) by linear expressions in $m^2_\pi$ and $a^2$.}
\label{fig:decay}
\end{figure}

\noindent
 We display in Fig.\ref{fig:decay} the extrapolations at the physical point of $f_{\eta_c}$ and $f_{J/\psi}$. They are mild, cut-off effects on $f_{\eta_c}$ are of the order of 4\% at $\beta=5.3$ while they are stronger for $f_{J/\psi}$, about 10\%. We get at the physical point
\begin{equation}\nonumber
f_{\eta_c}=387(3)(3)\, {\rm MeV}\quad\text{and} \quad f_{J/\psi}=399(4)(2)\, {\rm MeV},
\end{equation}
where the systematic error comes from the uncertainty on lattice spacings.
\newline
Moreover, one can derive a phenomenological estimate of $f_{J/\psi}$. Indeed, using the expression of the electronic decay width
\beq\nonumber
\Gamma(J/\psi \to e^+e^-)=\dfrac{4\pi}{3}\,\dfrac{4}{9}\,\alpha(m^2_c)\dfrac{f^2_{J/\psi}}{m^2_{J/\psi}}
\eeq
together with the experimental determination of the $J/\psi$ mass and width, and setting $\alpha_{\rm em}(m^2_c)={1}/{134}$~\cite{PivovarovCR}, one gets
$$f^{\text{``exp''}}_{J/\psi}=407(6)\,\text{MeV}$$
\par
\begin{figure}[ht] 
\begin{center}
\begin{tabular}{cc}
	\includegraphics[width=7.7cm,clip]{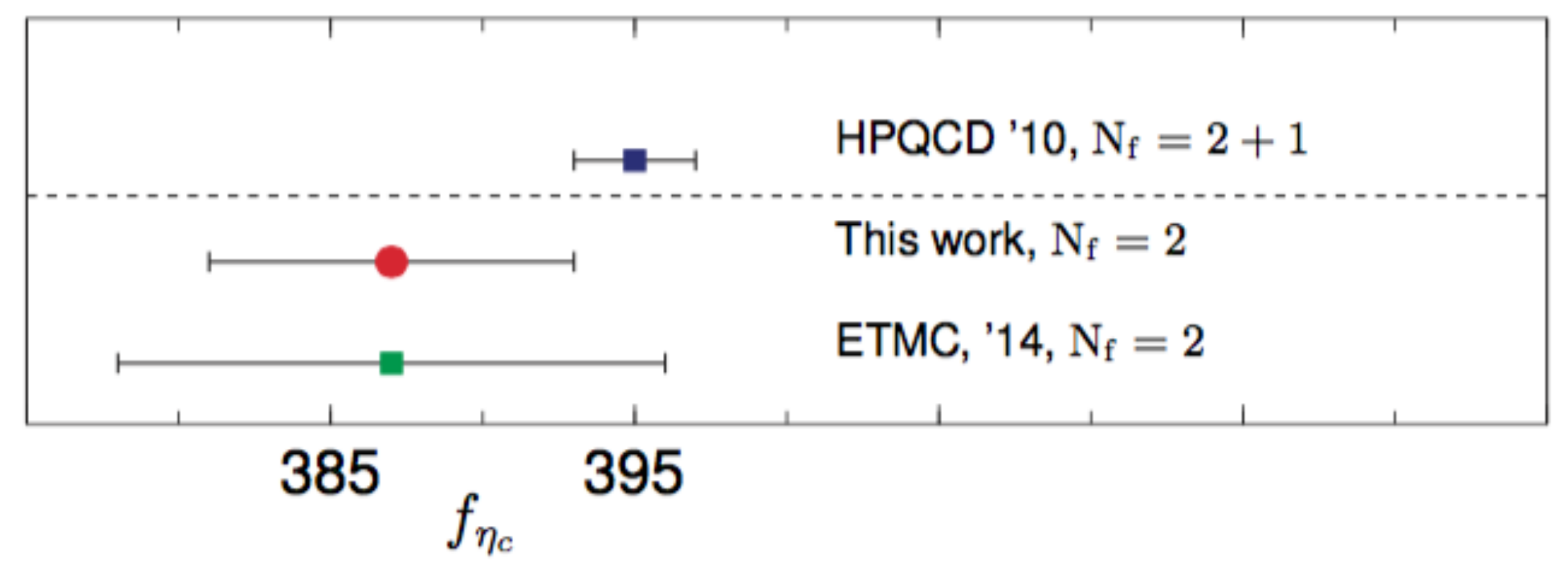}
&
	\includegraphics[width=7cm,clip]{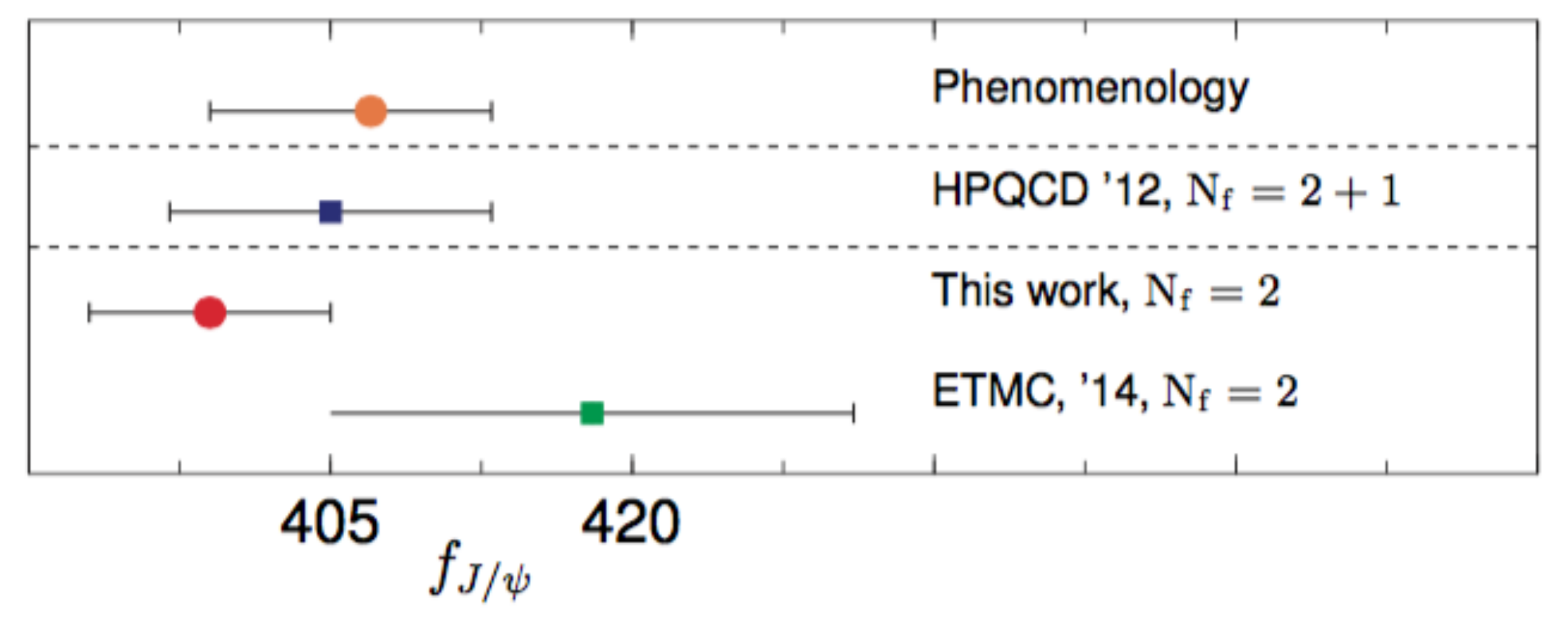}
\end{tabular}
\end{center}
	\caption{Collection of lattice results of $f_{\eta_c}$ (left panel) and $f_{J/\psi}$ (right panel).}
\label{fig:collectfetacfjpsi}
\end{figure}

\noindent 
We collect the various lattice QCD results and the phenomenological estimate of $f_{J/\psi}$ in Fig.\ref{fig:collectfetacfjpsi}. 
\par
\begin{figure}[t] 
	\begin{minipage}[c]{0.45\linewidth}
	\centering 
\includegraphics*[width=7cm, height=5cm]{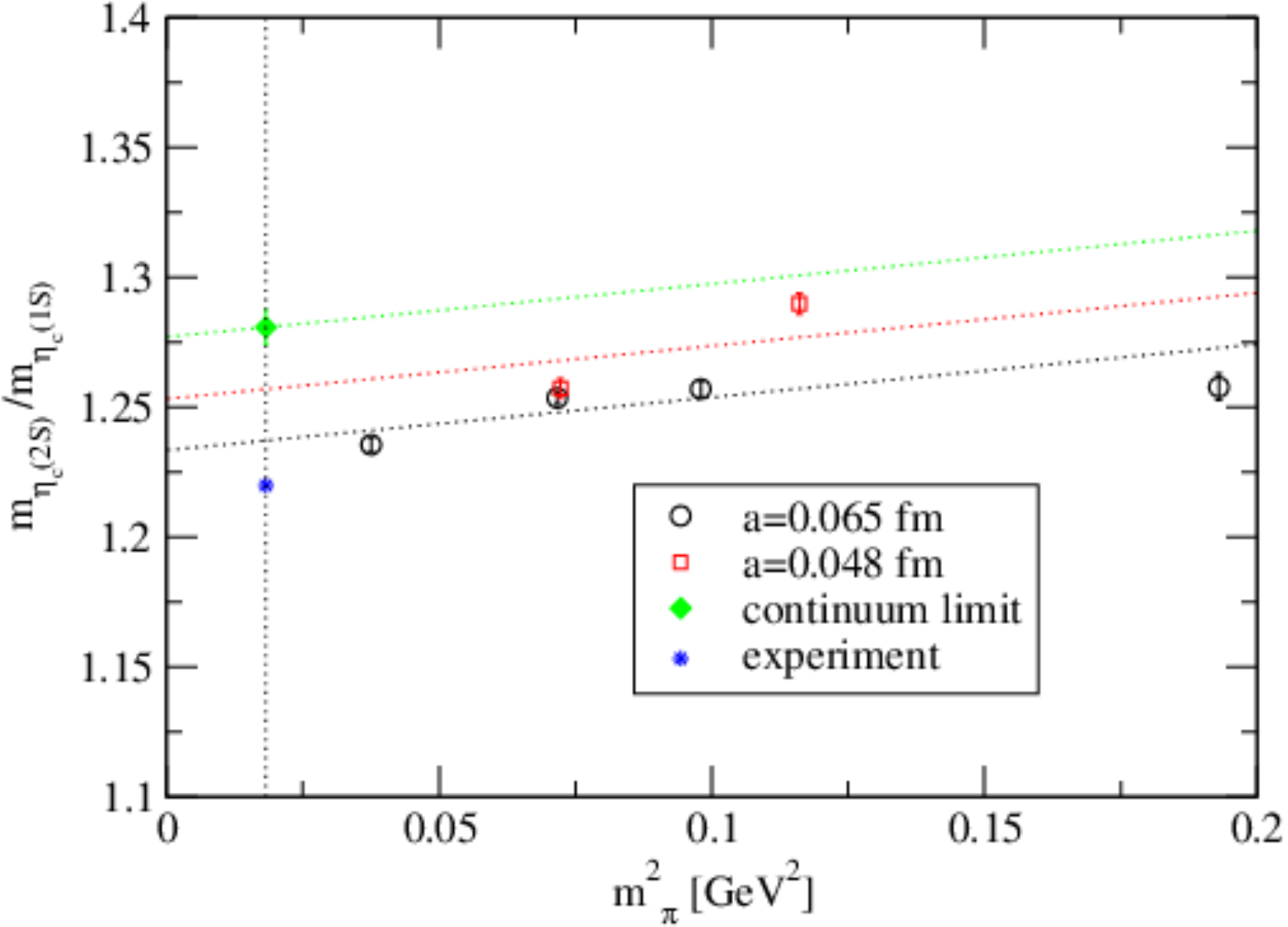}
	\end{minipage}
	\begin{minipage}[c]{0.45\linewidth}
	\centering 
\includegraphics*[width=7cm, height=5cm]{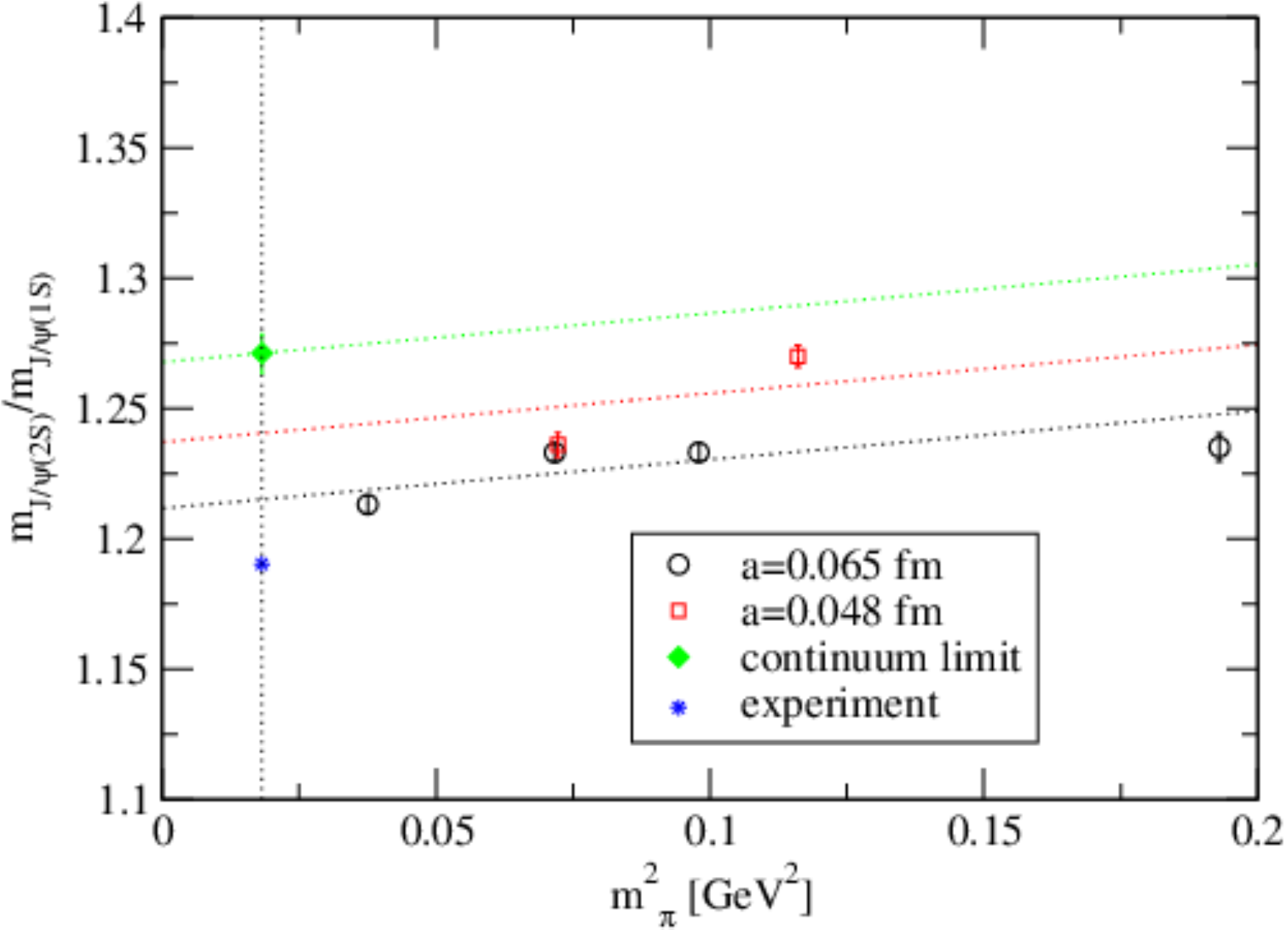}
	\end{minipage}
	\caption{Extrapolation at the physical point of $m_{\eta_c(2S)}/m_{\eta_c}$ (left panel) and $m_{\psi(2S)}/m_{J/\psi}$ (right panel) by linear expressions in $m^2_\pi$ and $a^2$.}
\label{fig:ratiomass}
\end{figure}

\noindent
Unfortunately the situation is less favorable with the excited states. Fig.\ref{fig:ratiomass} shows the extrapolation to the continuum limit
of the ratios $m_{\eta_c(2S)}/m_{\eta_c}$ and $m_{\psi(2S)}/m_{J/\psi}$, compared to the experimental values. Since the cut-off effects
are small, of the order of  5\%, there is no hope to have points in the continuum limit significantly lower than our lattice data. We get 
$$m_{\eta_c(2S)}/m_{\eta_c}=1.281(7)\ \gg\ (m_{\eta_c(2S)}/m_{\eta_c})^{\rm exp}=1.220$$ and $$m_{\psi(2S)}/m_{J/\psi}=1.271(7)\ \gg\ (m_{\psi(2S)}/m_{J/\psi})^{\rm exp}=1.190$$
In \cite{BecirevicRDA}, lattice results were much closer to the experiment but the slope in $a^2$ was probably overrestimated from the coarsest lattice point: points at lattice spacings similar to those used in our work are compatible with our data.\par
\begin{figure}[t] 
	\begin{minipage}[c]{0.45\linewidth}
	\centering 
\includegraphics*[width=7cm, height=5cm]{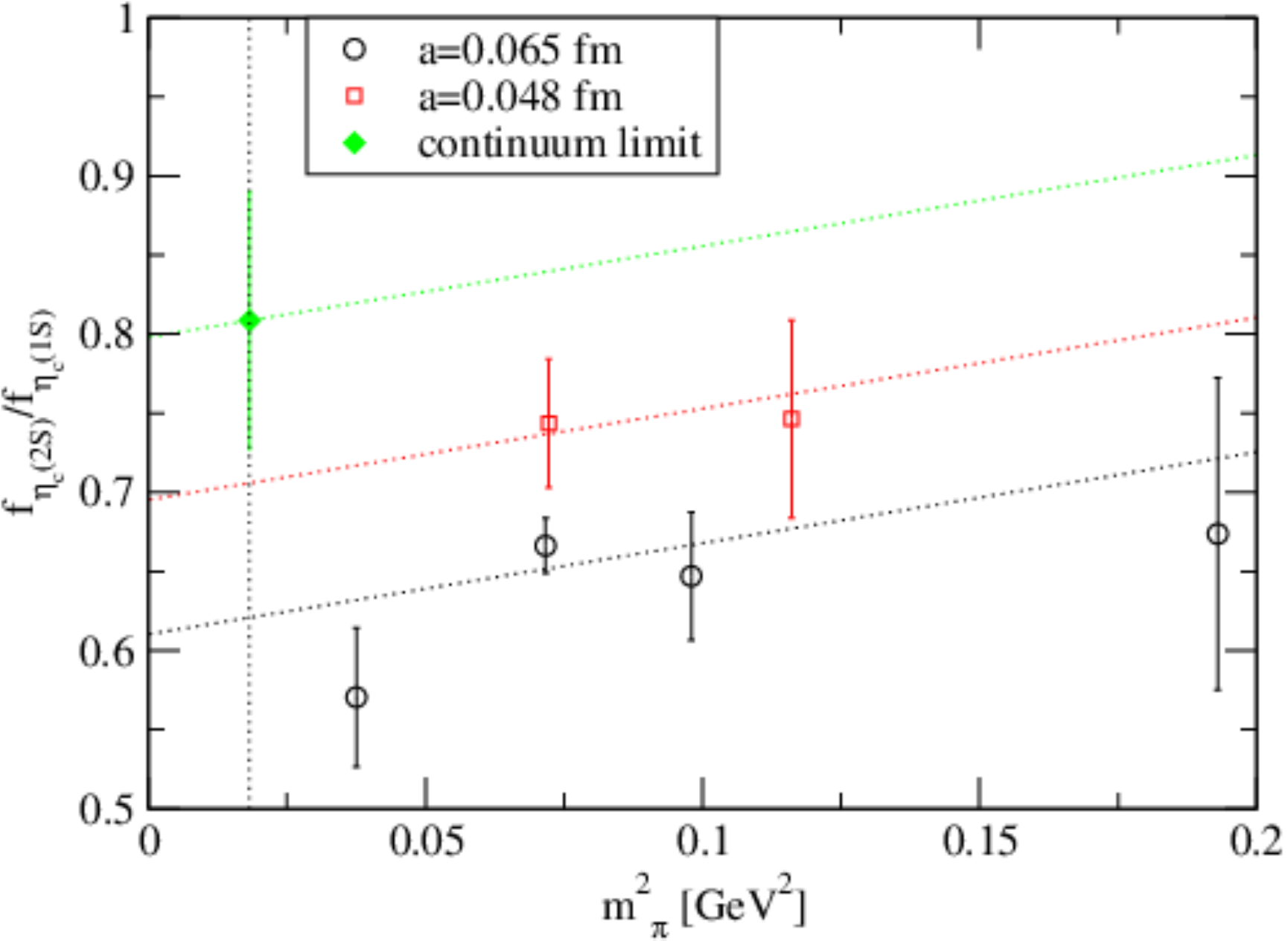}
	\end{minipage}
	\begin{minipage}[c]{0.45\linewidth}
	\centering 
\includegraphics*[width=7cm, height=5cm]{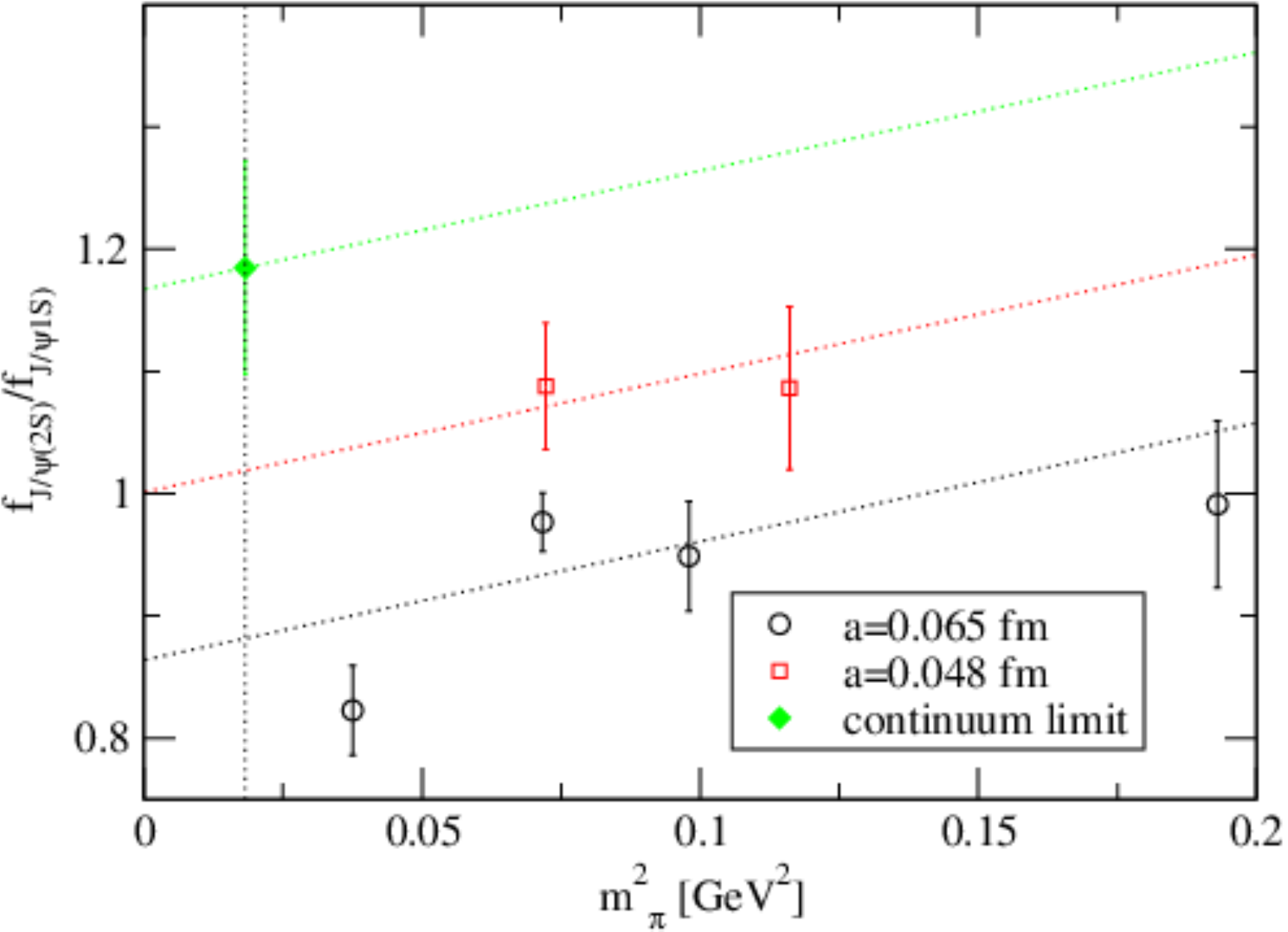}
	\end{minipage}
	\caption{Extrapolation at the physical point of $f_{\eta_c(2S)}/f_{\eta_c}$ (left panel) and $f_{\psi(2S)}/f_{J/\psi}$ (right panel) by linear expressions in $m^2_\pi$ and $a^2$.}
\label{fig:ratiodecay}
\end{figure}

\noindent
The situation is even more 
confusing for the ratios of decay constants, as shown in Fig.\ref{fig:ratiodecay} $$f_{\eta_c(2S)}/f_{\eta_c} =0.81(8)<1\qquad\text{while}\qquad f_{\psi(2S)}/f_{J/\psi} =1,18(9)>1$$
We have not found any explanation for such a large spin breaking effect. Projected correlators $\tilde{C}^2$ in the pseudoscalar and the vector sector show the same quality fit with similar fluctuations. We also have performed a global fit of individual correlators $C_{V^{(i)}V^{(j)}}(t)$ by a series of 3 exponential contributions
\begin{multline*}
C_{V^{(i)}V^{(j)}}(t)\ =\ Z^i_1Z^{*j}_1 e^{-m_{V_1} T/2} \cosh[m_{V_1}(T/2-t)]\\[1mm]
\ +\ Z^i_2Z^{*j}_2 e^{-m_{V_2} T/2} \cosh[m_{V_2}(T/2-t)]
\ +\ Z^{ij}_3 e^{-m^{ij}_3 T/2} \cosh[m^{ij}_3(T/2-t)]
\end{multline*}
where the ``effective'' remaining mass $m_3$, resumming any contributions but the ground state and the first excited state, can be different for every correlator. Decay constants of $J/\psi$ and $\psi(2S)$ are proportional to $Z^0_1/\sqrt{m_{V_1}}$ and $Z^0_2/\sqrt{m_{V_2}}$ (the local-local vector 2-pt correlator corresponds to $i\!=\!j\!=0$) and we find the hierarchy $Z^0_2/\sqrt{m_{V_2}}\gtrsim Z^0_1/\sqrt{m_{V_1}}$ in our sets of data. There is no hope to get in the continuum limit $f_{\psi(2S)}/f_{J/\psi}<1$ using this procedure either.
Neglecting disconnected diagrams might be a source of the problem; including more flavours in the sea might help to reduce the spin breaking effects, as it was observed on quantities like $f_{D^*_s}/f_{D_s}$ \cite{BecirevicTI, BlossierJOL, DonaldSRA, LubiczASP}. Besides, the width $\Gamma(\psi(2S) \to e^+e^-)$ is smaller than its ground state counterpart $\Gamma(J/\psi \to e^+e^-)$, that is 2.34 keV versus 5.56 keV~\cite{PatrignaniXQP}. Written in terms of the decay constant $f_{\psi(2S)}$ and the mass $m_{\psi(2S)}$, this is a serious clue that $f_{\psi(2S)}/f_{J/\psi}<1$ \footnote{A possible caveat with this approach is that QED effects might be quite large, making, as is done in our work, the encoding in $f_{\psi(2S)}$ as purely QCD contributions not so straightforward.}. Moreover, a lattice study, performed in the framework of NRQCD, gives $f_{\Upsilon'}/f_{\Upsilon}<1$ in the bottomium sector~\cite{ColquhounICA}.


\section{Conclusion}

In that paper we have reported a $N_f=2$ lattice QCD study about the physics of quarkonia. The decay constants $f_{\eta_c}$ and $f_{J/\psi}$ are in the same ballpark as the two previous lattice estimations available so far in the literature, with the good news that cut-off effects seem to be limited to 10\%. The issues
 with radial excitations are difficult to circumvent. As a first solution, the basis of operators in the GEVP analysis could be enlarged by including interpolating
fields with covariant derivatives or operators of the $\pi_2$ and $\rho_2$ kind. But both of them suffer either from big statistical fluctuations, because of numerical cancellation among various contributions, or from the more serious conceptual problem that, in GEVP, mixing T-even and T-odd operators has no real sense. Second, $m_{\eta_c(2S)}/m_{\eta_c}$ and $m_{\psi(2S)}/m_{J/\psi}$ are not significantly affected by cut-off effects, so that they are 5\% larger than the experimental ratios. The information that $f_{\eta_c(2s)}/f_{\eta_c} \sim 0.8$ is interesting, as the decay constants $f_{\eta_c}$ and $f_{\eta_c(2S)}$ are hadronic inputs that govern the transitions $\eta_c \to l^+ l^-$, $h \to \eta_c l^+ l^-$, $\eta_c(2S) \to l^+ l^-$ and $h \to \eta_c(2S) l^+ l^-$ with a light CP-odd Higgs boson as an intermediate state\footnote{In the cases $h \to \eta_c l^+ l^-$ and $h \to \eta_c(2S) l^+ l^-$, the other hadronic quantities which enter the process are the distribution amplitudes of the charmonia.}. Unfortunately, our result for $f_{\psi(2S)}/f_{J/\psi}>1$ makes the picture less bright, unless one admits that there are very large spin breaking effects. Further investigation to address this issue is undoubtedly required.\par\noindent
Nevertheless, the next step into the measurement of $f_{\eta_b}$, particularly relevant in models with a light CP-odd Higgs, is underway using step scaling in masses in order to extrapolate the results to the bottom region.

\section*{Acknowledgments} This work was granted access to the HPC resources of CINES and IDRIS under the allocations 2016-x2016056808 
and 2017-A0010506808 made by GENCI. Authors are grateful to Damir Becirevic and Olivier P\`ene for useful discussions and the colleagues of the CLS effort for having provided the gauge ensembles used in that work.

\section*{Appendix}

We collect in Table \ref{tabresults} the values of $\eta_c$ and $J/\psi$ masses and decay constants extracted at each ensemble of our analysis,
as well as the ratios of masses and decay constants $m_{\eta_c(2S)}/m_{\eta_c}$, $m_{\psi(2S)}/m_{J/\psi}$,
$f_{\eta_c(2S)}/f_{\eta_c}$ and $f_{\psi(2S)}/f_{J/\psi}$.

\begin{table}[t]
\begin{center}
\begin{tabular}{|c|c|c|c|c|c|c|}
\hline
id&$[t_{\rm min}, t_{\rm max}](P)$&$am_{\eta_c}$&$af_{\eta_c}$&$[t_{\rm min}, t_{\rm max}](V)$&$am_{J/\psi}$&$af_{J/\psi}$\\
\hline
E5&[11-29]&0.9836(3)&0.1246(16)&[11-29]&1.0202(7)&0.1499(11)\\
F6&[11-46]&0.9870(1)&0.1236(5)&[11-46]&1.0233(4)&0.1471(9)\\
F7&[11-45]&0.9855(1)&0.1233(3)&[11-45]&1.0209(3)&0.1460(5)\\
G8&[12-55]&0.9861(1)&0.1231(3)&[12-55]&1.0217(2)&0.1454(5)\\
N6&[13-46]&0.7284(3)&0.0944(6)&[13-46]&0.7547(6)&0.1059(8)\\
O7&[16-55]&0.7297(1)&0.0927(3)&[16-55]&0.7555(3)&0.1037(4)\\
\hline
\end{tabular}
$$\quad$$
\begin{tabular}{|c|c|c|c|c|c|c|}
\hline
id&$[t_{\rm min}, t_{\rm max}](P')$&$m_{\eta_c(2S)/m_{\eta_c}}$&$f_{\eta_c(2S)}/f_{\eta_c}$&$[t_{\rm min}, t_{\rm max}](V')$&
$m_{\psi(2S)}/m_{J/\psi}$&$f_{\psi(2S)}/f_{J/\psi}$\\
\hline
E5&[6-13]&1.258(5)&0.67(10)&[6-13]&1.235(5)&0.99(6)\\
F6&[6-13]&1.257(3)&0.65(4)&[6-13]&1.233(4)&0.95(4)\\
F7&[6-13]&1.254(2)&0.67(2)&[6-13]&1.233(3)&0.98(2)\\
G8&[8-15]&1.235(3)&0.57(4)&[8-15]&1.213(3)&0.82(4)\\
N6&[8-15]&1.290(4)&0.75(6)&[8-15]&1.270(4)&1.09(7)\\
O7&[8-15]&1.257(4)&0.74(4)&[8-15]&1.236(5)&1.09(5)\\
\hline
\end{tabular}
\end{center}
\caption{Masses and decays constants of $\eta_c$, $\eta_c(2S)$, $J/\psi$ and $\psi(2S)$, in lattice units, extracted 
on each CLS ensemble used in our analysis. \label{tabresults}}
\end{table}

\end{document}